\documentclass[12pts]{article}       
\usepackage{graphicx}
\begin{document}

\begin{center}
{\bf \Large Optimal decision for the market graph identification problem in sign similarity network.} \\
\end{center}
\begin{center}
Kalyagin V.A., Koldanov P.A., Pardalos P.M.
\footnote{National Research University Higher School of Economics, Laboratory of Algorithms and Technologies for Network Analysis, 
Bolshaya Pecherskaya 25/12, Nizhny Novgorod, RUSSIA. Tel.: +7-831-4361397. Fax: +7-831-4169655. Email: vkalyagin@hse.ru}  
\end{center}







\noindent
{\bf Abstract:}
Investigation of the market graph attracts a growing attention in market network analysis. One of the important problem connected with market graph is  to identify it from observations. 
Traditional way for the market graph identification is  to use a simple procedure  based on statistical estimations of  Pearson correlations between pairs of stocks. 
Recently a new class of statistical procedures for the market graph identification was introduced and optimality of these procedures  in Pearson correlation Gaussian network was proved.
However the  obtained  procedures have a high reliability only for Gaussian multivariate distributions of stocks attributes. One of the way to correct this drawback is  to consider 
a different networks generated by different  measures of pairwise similarity of stocks. A new and promising model in this context is the sign similarity network. 
In the present paper the market graph identification problem in sign similarity network is considered.  A new class of statistical procedures for the market graph identification is introduced 
and optimality of these procedures  is proved. Numerical experiments detect essential difference in quality of  optimal procedures  in sign similarity  and 
Pearson correlation networks. In particular it is observed that the quality of optimal identification procedure in sign similarity network is not sensitive to the  assumptions on  distribution
of stocks attributes.   


\section{Introduction}

There are a variety of data mining techniques applied for stock markets. Some of them are based on the analysis of market network and its structures. Market network is a complete weighted graph
where the nodes are associated with stocks and weights of edges are given by some measure of similarity between stocks behavior. {\it Market graph} is an important structure in market network. 
An edge between two nodes  is included in the market graph, iff the corresponding measure of similarity  is larger than a given threshold. Maximum cliques, 
maximum independent sets, degree distribution in the market graph are useful sources of market data mining. 

The concept of the market graph was introduced in \cite{Pardalos_1}. Since, different aspects
of the market graph approach (threshold method) were developed in the literature.  Most publications are related with experimental study of real markets. 
 The power law phenomena first observed for US stock market in \cite{Pardalos_2} was then developed in \cite{Huang}, \cite{Tse}, \cite{Pardalos_4}. 
Clustering in Pearson correlation based financial network is investigated in \cite{Onella_clustering}. Dynamics of the US  market graphs was studied in \cite{Pardalos_3}.
Complexity of the US market graph associated with significant correlations is investigated in \cite{Emmert_1}. Peculiarity of different financial markets are emphasized
in \cite{BKKKP}, \cite{Garas}, \cite{Vizgunov}, \cite{Huang}, \cite{Namaki}.  Market graphs with different measures of similarity were investigated in 
\cite{BKK13}, \cite{BKKKP}, \cite{Hero}, \cite{Shirokikh}, \cite{DTW}, \cite{Partial}.
Some efficient algorithms related with calculation of isolated cliques in a market graph are presented in \cite{Gunawardena}, \cite{Huffner}.

However, economical interpretation of market network data mining is not complete without estimation of reliability  of obtained results.   
Reliability of minimum spanning tree in Pearson correlation based network is investigated by bootstrap method in  \cite{Tumminello_reliability}. 
In the present paper we use a different approach to handle reliability of the market graph. Our approach is based on the model of {\it random variables network}.   
The nodes of the network are random variables and weights of edges are given by some measure of pairwise similarity between the random variables. 
Observed values of stocks attributes are modeled by sample from distribution of random variables. 
Reliability  of network structure can now be measured by risk function of statistical procedure for its identification. Statistical procedures with minimal risk (maximal reliability) are of grate
practical interest.  Class of optimal statistical procedures for the identification of market graph in Pearson correlation based network were introduced and investigated  in  \cite{KKKP13}.
One can see from numerical experiments in \cite{KKKP13} that  the value of the risk function of the procedures of this class essentially depends on the assumptions on multivariate distributions of stocks attributes. 
Taking this into account it is of interest to investigate a distribution free identification statistical procedures. A new and promising approach in this context is to consider a sign correlation based (sign similarity) network. 

In this paper we investigate optimal statistical procedures for the market graph identification in sign similarity (sign correlation based) network.  
Our construction of optimal procedures is based on simultaneous inference of optimal two-decision procedures. 
It is proved that constructed procedure is optimal under the following assumptions: additivity of loss function, unbiasedness of procedure, sign symmetry of distributions.  
We give a direct proof which simplify a general approach  by Lehmann \cite{Lehmann}. In addition we compare the risk function of the optimal procedure  in sign similarity network with the 
risk function  of the optimal procedure in Pearson correlation network.  Numerical experiments detect essential difference in risk behavior for two optimal statistical procedures. 
For multivariate Gaussian distribution both procedures control the risk with a change of significance level of individual tests. In contrast for multivariate Student distribution optimal procedure in 
Pearson correlation network does not control the risk while the optimal procedure in sign similarity network does. It means that the quality of optimal identification procedure in sign similarity network 
is not sensitive to the  assumptions on  distribution of stocks attributes.

The paper is organized as follows. In the section \ref{Problem statement} we give a basic definitions and notations. 
In  section \ref{Multiple decision framework} we describe a multiple decision framework for threshold graph identification problem. 
In Section \ref{Concept of optimality} we discuss a concept of optimality of  identification procedures. 
In Section \ref{Multiple_decision_procedure} we construct a  multiple decision identification procedure in sign similarity network. 
In Section \ref{Optimality of multiple decision procedure} we give a proof of optimality of this procedure. 
In section \ref{Comparison of optimal procedures} we conduct a numerical experiments to compare optimal procedures in sign similarity and Pearson correlation network. 
In section \ref{Conclusions} we present a concluding remarks.

\section{Market graph identification problem}\label{Problem statement} 

Consider a network generated by a random vector $X=(X_1, X_2,\ldots, X_N)$.
Nodes of network are random variables $X_i$, $i=1,\ldots,N$ and weight of edge $(i,j)$ 
is given by some pairwise measure of association $\gamma$: 
$$
\gamma_{i,j}=\gamma(X_i,X_j),  \mbox{ for } i,j = 1,2, \ldots, N.
$$ 
The obtained network is a complete weighted graph which we will call {\it random variables network}. 
The random variables network is defined  by multivariate distribution of the vector $X$ and by the choice of measure of association $\gamma$. 
The network based on Pearson correlation of random variables will be called {\it Pearson correlation network}.
The network based on probability  of  pairwise sign coincidence will be called {\it sign similarity network}. 

For any network the market graph is constructed as follows: the edge between two vertices
\emph{i} and \emph{j} is included in the market graph, iff
$\gamma_{i,j} > \gamma_0$, where $\gamma_0$ is a given threshold. In what follows we will call this network structure {\it reference (true) market graph}.

In sign similarity network weight of edge
$(i,j)$ is defined by
\begin{equation}\label{Fechner measure}
p^{i,j}=P((X_i-E(X_i))(X_j-E(X_j))>0)   
\end{equation}
For a given threshold $p_0$ {\it reference market graph} in sign similarity network is constructed as follows: 
edge between two nodes $i$ and $j$ is included in the reference market graph iff $p^{i,j}>p_0$, where $p^{i,j}$ is the probability of sign coincidence of random variables  associated with nodes $i$ and $j$.

In practice $\gamma_{i,j}$ are unknown and we are given a sample of observations $x(1), x(2), \ldots, x(t)$ from distribution $X$. Identification  of reference market graph 
from observations is called in this paper {\it market graph identification problem}. 

Two types of errors are possible for any identification procedure.  Type I error occurs if identification procedures includes edge in the market graph when it is absent in the reference market graph. Type II error occurs if identification procedures does not include edge in the market graph when it is present in the reference market  graph. For market graph identification it is important to control not only type I and type II errors but the number of errors. 


\section{Multiple decision framework}\label{Multiple decision framework}
We model observations as a family of random vectors 
$$
X(t)=(X_1(t),X_2(t),\ldots,X_N(t)), \quad t =1, 2,\ldots,n
$$
where $n$ is the number of observations (sample size) and vectors $X(t)$ are independent and identically distributed as $X=(X_1,X_2,\ldots,X_N)$.
In what follows we assume that expectations $E(X_i)$, $i=1,2,\ldots, N$ are known. We put (for simplicity)  $E(X_i)=0$, $i=1,2,\ldots,N$. In this case
\begin{equation}\label{p_i_j}
p^{i,j}=P(X_iX_j>0), \ \ i,j=1,2,\ldots,N
\end{equation}
The random vector $X$ with measures of association $(p^{i,j})$ define a sign similarity network. For a given threshold $p_0$ the reference market graph
is defined by its adjacency matrix $TG=(tg_{i,j})$, where $tg_{i,j}=0$ if $p^{i,j} \leq p_0$ and $tg_{i,j}=1$ if $p^{i,j} > p_0$, $tg_{i,i}=0$, $i,j=1,2,\ldots,N$. 

Let $x_{i}(t)$ be  observations of the random variables $X_i(t)$, $t =1, 2, \ldots,n$, $i = 1,2, \ldots,N$. 
Consider the set $\cal{G}$ of all $N \times N$ symmetric matrices  $G=(g_{i,j})$ with $g_{i,j} \in \{0,1\}$, $i,j=1,2,\ldots,N$, $g_{i,i}=0$, $i=1,2,\ldots,N$. Matrices $G \in \cal{G}$ represent adjacency 
matrices of all simple undirected graphs with $N$ vertices. The total number of matrices in $\cal{G}$ is equal to $L=2^M$ with $M=N(N-1)/2$. 
The {\it market graph identification problem} in sign similarity network 
can be formulated as a multiple decision problem of the selection of one from the set of $L$ hypotheses:
\begin{equation}\label{N hypotheses}
H_G: p^{i,j}\leq p_0,\mbox{ if } g_{i,j}=0, \ \ p^{i,j}>p_0,\mbox{ if } g_{i,j}=1; \ \ i \neq j
\end{equation}

Consider some examples of hypotheses (\ref{N hypotheses}).  For the matrix $$G_1=\left(\begin{array}{cccc}
0&0&\ldots&0\\
0&0&\ldots&0\\
\ldots&\ldots&\ldots&\ldots\\
0&0&\ldots&0\\
\end{array}\right)$$
the corresponding graph has $N$ isolated vertices and the associated hypothesis $H_{G_1}$ is
$$H_{G_1}:p^{i,j}\leq p_0,\forall i,j=1,\ldots,N, \ \ i \neq j$$
 
For the  matrix:
$$G_2=\left(\begin{array}{ccccc}
0&1,&0&\ldots&0\\
1&0&0&\ldots&0\\
0&0&0&\ldots&0\\
\ldots&\ldots&\ldots&\ldots\\
0&0&0&\ldots&0\\
\end{array}\right)$$
the corresponding graph has only one edge $(1,2)$ and the associated hypothesis $H_{G_2}$ is 
$$H_{G_2}:p^{1,2}>p_0, p^{2,1}>p_0, p^{i,j}\leq p_0,\forall (i,j)\neq(1,2), (i,j)\neq(2,1), i \neq j$$

For the matrix:
$$G_3=\left(\begin{array}{cccc}
0&1&\ldots&1\\
1&0&\ldots&1\\
\ldots&\ldots&\ldots&\ldots\\
1&1&\ldots&0\\
\end{array}\right).$$
the corresponding graph is a complete graph and the associated hypothesis $H_{G_3}$ is
$$H_{G_3}:p^{i,j} > p_0,\forall i,j=1,\ldots,N, \ \ i \neq j$$

To solve the identification problem (\ref{N hypotheses}) one has to use multiple decision statistical procedures. Multiple decision statistical procedure $\delta$ 
is a map from the sample space $R^{N \times n}$ to the decision space $D=\{d_G, g \in \cal{G} \}$,  where the decision  $d_G$ is the
acceptance of hypothesis $H_G$, $G \in \cal{G}$. 

\section{Concept of optimality}\label{Concept of optimality}
In this section we discuss a concept of optimality related with multiple decision statistical procedures. According to \cite{Wald} the quality of statistical procedure 
is defined by risk function. Consider a statistical procedure $\delta(x)$. 
Let $S=(s_{i,j})$, $Q=(q_{i,j})$, $S,Q \in \cal{G}$. Denote by $w(S,Q)$ the loss from the decision $d_Q$ when the hypothesis $H_S$ is true
$$w(H_S;d_Q)=w(S,Q), \ \ S,Q \in \cal{G}$$ 
It is assumed that $w(S,S)=0, S \in \cal{G}$.
Risk function $Risk: \cal{G} \to R$ is defined by
$$
Risk(S;\delta)=\sum_{Q \in \cal{G}} w(S,Q)P(\delta(x)=d_Q/H_S), \quad S \in \cal{G}
$$
where $P(\delta(x)=d_Q/H_S)$ is the probability that decision $d_Q$ is taken while the true decision is $d_S$. 
The problem of minimization of $Risk$ is therefore multiple criteria decision problem: optimal procedure $\delta$ has to minimize $Risk(S,\delta)$
for every $S \in \cal{G}$. In general such problem does not have a solution. Instead  one can use a Pareto optimal solutions. 
However, it is possible to get a solution if one imposes a constraints on the procedures. One common constraint is unbiasedness of the procedure. Following \cite{Lehmann_Romano}
we call the multiple decision procedure $\delta(x)$ $w$-unbiased if 
\begin{equation}\label{w-unbiasedness}
\sum_{Q \in \cal{G}} w(S,Q)P(\delta(x)=d_Q/H_S) \leq \sum_{Q \in \cal{G}} w(S',Q)P(\delta(x)=d_Q/H_S), \ \ \forall S,S' \in \cal{G}
\end{equation}
``Thus $\delta$ is unbiased if on the average $\delta(x)$ comes closer to the correct decision than to any wrong one'' (citation from \cite{Lehmann_Romano}, page 13). 
Note that unbiasedness depends on the  loss function $w$. 

For the market graph identification problem it is important to control not only type I and type II errors but a number of errors. 
Let $a_{i,j}$ be the loss from false inclusion of edge $(i,j)$ in market graph and let $b_{i,j}$, be the
loss from false non inclusion of the edge $(i,j)$ in the market graph, $i,j=1,2,\ldots,N; \ i\neq j$.  

Let
$$l_{i,j}(S,Q)=\left\{\begin{array}{cc}
a_{i,j}, & \mbox{if } \ s_{i,j}=0,q_{i,j}=1, \\
b_{i,j}, & \mbox{if } \ s_{i,j}=1,q_{i,j}=0, \\
0, & \mbox{ else }
\end{array}\right.$$  

It is natural to define loss function $w(S,Q)$ as:
\begin{equation}\label{additive_loss_function}
w(S,Q)=\sum_{i=1}^N\sum_{j=1}^N l_{i,j} (S,Q)
\end{equation}
It means that the loss from misclassification of $H_S$ is equal to the sum of losses from misclassification of individual edges:
$$
w(S,Q)=\sum_{\{i,j:s_{i,j}=0;q_{i,j}=1\}}a_{i,j}+\sum_{\{i,j:s_{i,j}=1;q_{i,j}=0\}}b_{i,j} 
$$
In the next sections we investigate the optimality of multiple statistical procedures for market graph identification for the loss function in the class of unbiased procedures.

\section{Multiple decision procedure based on simultaneous inference of two decision tests}\label{Multiple_decision_procedure}
In this section we describe a class of multiple decision procedures based on simultaneous inference of individual edge tests. 

Any individual edge test can be reduced to hypotheses testing problem: $h_{ij}:\gamma_{ij}\leq\gamma_0$ vs $k_{ij}:\gamma_{ij}>\gamma_0$.
According to \cite{Lehmann_Romano} test of the individual edge hypotheses has the form:
\begin{equation}\label{ind_edge_test_overall_form}
\varphi_{ij}(x)=\left\{ \begin{array}{cc} 0,& t_{ij}(x)\leq c_{\alpha}\\1,&t_{ij}(x)> c_{\alpha} \end{array}\right.
\end{equation}  
where $\varphi_{ij}(x)=1$ means that edge $(i,j)$ is included in the market graph, $\varphi_{ij}(x)=0$ means that edge $(i,j)$ does not included in the market graph. 
 
Let $\Phi(x)$ be the matrix
\begin{equation}\label{test_for_N_hypotheses_overall_form}
\Phi(x)=\left(\begin{array}{cccc}
1,&\varphi_{12}(x),&\ldots,&\varphi_{1N}(x)\\
\varphi_{21}(x),&1,&\ldots,&\varphi_{2N}(x)\\
\ldots&\ldots&\ldots&\ldots\\
\varphi_{N1}(x),&\varphi_{N2}(x),&\ldots,&1\\
\end{array}\right).
\end{equation}
Then all statistical procedures from the class of multiple decision procedures based on simultaneous inference of individual edge tests can be written as
\begin{equation}\label{mdp_overall_form}
\delta(x)=d_G, \  \mbox{iff} \  \Phi(x)=G
\end{equation}

Consider the sign similarity network.
Let
$$p^{i,j}_{0,0}=P(X_i\leq 0, X_j\leq0), \quad p^{i,j}_{1,1}=P(X_i>0, X_j>0)$$ 
$$p^{i,j}_{1,0}=P(X_i>0, X_j\leq0), \quad p^{i,j}_{0,1}=P(X_i\leq0, X_j>0)$$
One has $p^{i,j}=p^{i,j}_{0,0}+p^{i,j}_{1,1}$. Define 
$$u_k(t)=\left\{ \begin{array}{cc}
					0,&x_k(t)\leq 0\\
					1,&x_k(t)>0
					\end{array}\right.$$
$k=1,2,\ldots,N$. Let us introduce statistics 
\begin{equation}\label{statistics_for_sign}
\begin{array}{cc}
T_{1,1}^{i,j}=\sum_{t=1}^n u_i(t)u_j(t); & T_{0,0}^{i,j}=\sum_{t=1}^n (1-u_i(t))(1-u_j(t));\\
T_{0,1}^{i,j}=\sum_{t=1}^n (1-u_i(t))u_j(t);& T_{1,0}^{i,j}=\sum_{t=1}^n u_i(t)(1-u_j(t)); \\
V_{i,j}=T_{1,1}^{i,j}+T_{0,0}^{i,j} &  \\
\end{array}
\end{equation}
To construct a multiple decision procedure we use the following individual edge tests:
\begin{equation}\label{optimal_sign_test}
\varphi_{i,j}^{Sg}(x_i,x_j)=\left\{\begin{array}{cc}
										0,&V_{i,j}\leq c_{i,j}\\
										1,&V_{i,j}>c_{i,j}
										\end{array}\right.
\end{equation}
where for a given significance level $\alpha_{i,j}$, the constant $c_{i,j}$ is defined as minimal entire number such that:
\begin{equation}\label{threshold_equation_binomial_distribution}
\sum_{k=c_{i,j}}^n\frac{n!}{k!(n-k)!}(p_0)^k(1-p_0)^{n-k}\leq\alpha_{i,j}
\end{equation}

Let $\Phi^{Sg}(x)$ be the matrix
\begin{equation}\label{test_for_N_hypotheses}
\Phi^{Sg}(x)=\left(\begin{array}{cccc}
1,&\varphi^{Sg}_{12}(x),&\ldots,&\varphi^{Sg}_{1N}(x)\\
\varphi^{Sg}_{21}(x),&1,&\ldots,&\varphi^{Sg}_{2N}(x)\\
\ldots&\ldots&\ldots&\ldots\\
\varphi^{Sg}_{N1}(x),&\varphi^{Sg}_{N2}(x),&\ldots,&1\\
\end{array}\right).
\end{equation}
where $\varphi^{Sg} _{ij}(x)$ are defined by (\ref{optimal_sign_test})-(\ref{threshold_equation_binomial_distribution}. 
Now we can define our multiple decision statistical procedure for market graph identification 
\begin{equation}\label{mdp_another_form}
\delta^{Sg}(x)=d_G, \  \mbox{iff} \  \Phi^{Sg}(x)=G
\end{equation}
Constructed procedure looks very natural for market graph identification in sign similarity network. 

\section{Optimality of decision procedure $\delta^{Sg}$.}\label{Optimality of multiple decision procedure}
In this section conditions of optimality of multiple decision statistical procedure defined by (\ref{optimal_sign_test})-(\ref{mdp_another_form}) among multiple decision statistical procedures based on two decision tests are described.
\newtheorem{teo}{Theorem}
\begin{teo}
Let loss function $w$ be given by (\ref{additive_loss_function}), individual test statistics $t_{ij}$(\ref{ind_edge_test_overall_form}) depends only on $u_i(t), u_j(t)$, and following symmetry conditions are satisfied
\begin{equation}\label{symmetry_conditions}
 p_{11}^{ij}=p_{00}^{ij}, \quad p_{10}^{ij}=p_{01}^{ij},\quad \forall i,j = 1,2,\ldots, N
\end{equation} 
Then for the statistical procedure $\delta^{Sg}$ defined by (\ref{optimal_sign_test})-(\ref{mdp_another_form}) for market graph identification in sign similarity network one has $Risk(S,\delta^{Sg})\leq Risk(S,\delta)$ for any adjacency matrix $S$ and any $w$-unbiased statistical procedure
$\delta$. 
\end{teo}

\noindent
{\bf Proof}
We prove optimality in three steps. First we prove that under symmetry conditions (\ref{symmetry_conditions}) each individual test (\ref{optimal_sign_test}) is uniformly most powerful (UMP) in the class of tests based on $u_i(t),u_j(t)$ only  
for individual hypothesis testing
\begin{equation}\label{individual_sign_hypotheses}
h_{i,j}: p^{i,j} \leq p_0 \ \ \mbox{vs} \ \ k_{i,j}: p^{i,j} > p_0
\end{equation}
By symmetry conditions individual hypothesis (\ref{individual_sign_hypotheses}) can be written as:
\begin{equation}\label{individual_sign_hypotheses_exponential_form}
h_{i,j}: p_{00}^{i,j} \leq \frac{p_0}{2} \ \ \mbox{vs} \ \ k_{i,j}: p_{00}^{i,j} > \frac{p_0}{2}
\end{equation}
Let  $p_{0,0}=p_{0,0}^{i,j}; \ p_{0,1}=p_{0,1}^{i,j}; \ p_{1,0}=p_{1,0}^{i,j}; \ p_{1,1}=p_{1,1}^{i,j}$, $T_{0,0}=T_{0,0}^{i,j}; \ T_{0,1}=T_{0,1}^{i,j}; \ T_{1,0}=T_{1,0}^{i,j}; \ T_{1,1}=T_{1,1}^{i,j}$. 
One has
$$T_{1,1}+T_{1,0}+T_{0,1}+T_{0,0}=n; $$
Symmetry condition implies 
$$p_{0,0}+p_{1,0}=\frac{1}{2}$$

Let 
$t_{1,1},t_{1,0},t_{0,1},t_{0,0}$ be a non negative entire numbers with 
$t_{1,1}+t_{1,0}+t_{0,1}+t_{0,0}=n$ and   $C=n!/(t_{1,1}!t_{1,0}!t_{0,1}!t_{0,0}!)$. One has  
$$P(T_{1,1}=t_{1,1};T_{1,0}=t_{1,0};T_{0,1}=t_{0,1};T_{0,0}=t_{0,0})=C p_{1,1}^{t_{1,1}}p_{1,0}^{t_{1,0}}p_{0,1}^{t_{0,1}}p_{0,0}^{t_{0,0}}=$$
$$=C p_{0,0}^{t_{1,1}+t_{0,0}}p_{1,0}^{t_{1,0}+t_{0,1}}=C_1 \exp\{(t_{1,1}+t_{0,0})\ln\frac{p_{0,0}}{1/2-p_{0,0}}\}$$
where $C_1=C(1/2-p_{0,0})^n$.

Then hypotheses (\ref{individual_sign_hypotheses_exponential_form}) are equivalent to the hypotheses:
\begin{equation}\label{real_generating_hypothesis_12}
h'_{i,j}:\ln(\frac{p_{0,0}}{1/2-p_{0,0}})\leq\ln(\frac{p_{0}}{1-p_{0}})\ \mbox{ vs } \ k'_{i,j}:\ln(\frac{p_{0,0}}{1/2-p_{0,0}})>\ln(\frac{p_{0}}{1-p_{0}})
\end{equation}

For $p_{0,0}=p_0/2$ random variable $V=T_{1,1}+T_{0,0}$ has the binomial distribution $B(n,p_0)$. Therefore, critical value for the test (\ref{optimal_sign_test}) is 
defined from (\ref{threshold_equation_binomial_distribution}). According to (\cite{Lehmann_Romano}, Ch.3, corollary 3.4.1) the test (\ref{optimal_sign_test}) is uniformly most powerful (UMP) at the level $\alpha_{i,j}$ for hypothesis testing (\ref{real_generating_hypothesis_12}).  
 
Second we prove that statistical procedure (\ref{mdp_another_form}) is ${\it w}$-unbiased. For any two-decision test for hypothesis testing (\ref{individual_sign_hypotheses}) the risk function can be written as:
$$
Risk=R(s_{i,j},\varphi_{i,j})=\left\{\begin{array}{lll}
a_{i,j}P(\varphi_{i,j}(x)=1/p^{i,j}), & \mbox{if} & s_{i,j}=0 (p^{i,j}\leq p_0) \\ 
b_{i,j}P(\varphi_{i,j}(x)=0/p^{i,j}), & \mbox{if} & s_{i,j}=1 (p^{i,j}>p_0) \\  
\end{array}\right.
$$
Note that by general principle UMP test (\ref{optimal_sign_test}) is w-unbiased (\cite{Lehmann_Romano}, Ch.4).
More precisely, one has 
$$a_{i,j}P(\varphi^{Sg}_{i,j}(x)=1/p^{i,j})\leq b_{i,j}P(\varphi^{Sg}_{i,j}(x)=0/p^{i,j})  \mbox{ if }  p_{i,j}\leq p_0 $$ 
$$a_{i,j}P(\varphi^{Sg}_{i,j}(x)=1/p^{i,j})\geq b_{i,j}P(\varphi^{Sg}_{i,j}(x)=0/p^{i,j}), \mbox{ if }  p_{i,j}>p_0 $$
which is equivalent to
\begin{equation}\label{cond_risk_for_opt}
R(s_{i,j},\varphi^{Sg}_{i,j})\leq R(s'_{i,j},\varphi^{Sg}_{i,j}),\forall s_{i,j},s'_{i,j}
\end{equation}
This relation implies 
$$P(\varphi^{Sg}_{i,j}(x)=1/p^{i,j}=p_0)=\alpha_{i,j}=\frac{b_{i,j}}{a_{i,j}+b_{i,j}}$$
It means that test $\varphi^{Sg}_{i,j}$ has the significance level $\alpha_{i,j}=b_{i,j}/(a_{i,j}+b_{i,j})$. 
For loss function (\ref{additive_loss_function}) and any multiple decision statistical procedure $\delta$ one has
\begin{equation}\label{risk_function_additive_loss_function}
\begin{array}{c}
R(H_S,\delta)=\sum_{Q \in \cal{G}}(\sum_{i,j:s_{i,j}=0;q_{i,j}=1}a_{i,j}+\sum_{i,j:s_{i,j}=1;q_{i,j}=0}b_{i,j})P(x\in D_Q/H_S)=\\
=\sum_{s_{i,j}=0}a_{i,j}P(\varphi_{i,j}(x)=1)/H_S)+\sum_{s_{i,j}=1}b_{i,j}P(\varphi_{i,j}(x)=0)/H_S)
\end{array}
\end{equation}
Therefore: 
\begin{equation}\label{CR}
R(H_S,\delta)=\sum_{i=1}^N\sum_{j=1}^N R(s_{i,j};\varphi_{i,j})
\end{equation}
From (\ref{cond_risk_for_opt}) one has 
\begin{equation}\label{risk_function_final_form}
\sum_{Q \in \cal{G}} w(S,Q)P(\delta^{Sg}(x)=d_Q/H_S) \leq \sum_{Q \in \cal{G}} w(S',Q)P(\delta^{Sg}(x)=d_Q/H_S), \ \ \forall S,S' \in \cal{G}
\end{equation}
This means that multiple testing statistical procedure $\delta^{Sg}$ is unbiased.
 
Third we prove that procedure (\ref{mdp_another_form}) is optimal in the class of unbiased statistical procedures for market graph identification in sign similarity network. 
Let $\delta(x)$ be another unbiased statistical procedure for market graph identification in sign similarity network. Then $\delta(x)$ generates a partition of sample space $R^{N \times n}$ on $L$ parts:
$$D_G=\{x\in R^{N \times n}:\delta(x)=G\};\bigcup_{G\in\cal{G}}D_G=R^{N\times n}$$ 
Define 
\begin{equation}\label{mdp_generating}
\begin{array}{c}
A_{i,j}=\bigcup_{G: g_{i,j}(x)=0} D_G \\         
\overline{A_{i,j}}=\bigcup_{G:  g_{i,j}(x)=1} D_G
\end{array}
\end{equation}
and 
\begin{equation}\label{individual_edge_notoptimal_test}
\varphi_{i,j}(x)=\left\{
\begin{array}{cc}
0,& x\in A_{i,j}\\
1,& x \notin A_{i,j}
\end{array}\right.
\end{equation}
Tests (\ref{individual_edge_notoptimal_test}) are tests for individual hypotheses testing (\ref{individual_sign_hypotheses}). 
Since procedure $\delta(x)$ is unbiased then one has
$$
\sum_{Q \in \cal{G}} w(S,Q)P(\delta(x)=d_Q/H_S) \leq \sum_{Q \in \cal{G}} w(S',Q)P(\delta(x)=d_Q/H_S), \ \ \forall S,S' \in \cal{G}
$$
Consider the hypotheses $H_S$ and $H_{S'}$  which are different only in two components $s_{i,j}\neq s'_{i,j};\ s_{j,i}\neq s'_{j,i}$. Taking into account the unbiasedness of procedure $\delta$ and structure 
of the loss function (\ref{additive_loss_function}) one has  $R(s_{i,j},\varphi_{i,j})\leq R(s'_{i,j},\varphi_{i,j})$. This means that two decision tests (\ref{individual_edge_notoptimal_test}) are unbiased. Therefore
$$
P(\varphi_{i,j}=1/p_0)=\alpha_{i,j}=\frac{b_{i,j}}{a_{i,j}+b_{i,j}}.
$$
Since we are restricted the tests based on $u_i(t), u_j(t)$ only and test $\varphi^{Sg}_{i,j}$ is UMP among tests of the class at the significance level $\alpha_{i,j}$ then for any test $\varphi_{i,j}$ based on   $u_i(t), u_j(t)$ only one has:
$$R(s_{i,j},\varphi^{Sg}_{i,j})\leq R(s_{i,j},\varphi_{i,j})$$ 
From (\ref{CR}) one has
$$R(H_S,\delta^{Sg})\leq R(H_S,\delta)$$
for any adjacency matrix $S$.
Optimality of multiple testing statistical procedure $\delta^{Sg}$ has been proved.

\section{Numerical comparison of optimal procedures in sign similarity and Pearson correlation networks}\label{Comparison of optimal procedures}
In this section we compare the behavior of risk functions for two optimal multiple decision procedures:  identification procedure in sign similarity network (\ref{optimal_sign_test})-(\ref{mdp_another_form})
and  identification procedure in Pearson correlation network with Gaussian distribution. Optimal identification procedure in Pearson correlation network with Gaussian distribution was investigated in
\cite{KKKP13}. This procedure can be presented as follows. Define sample Pearson correlations 
$$
r_{ij}=\frac{\sum_{t}x_i(t)x_j(t)}{\sqrt{\sum_tx_i(t)^2\sum_t x_j(t)^2}}
$$
Let $\rho_0$ be the threshold. We use the following individual edge tests:
\begin{equation}\label{Pearson_test}
\varphi^{P}_{i,j}(x_i,x_j)=\left\{\begin{array}{cc}
										0,&z_{i,j}\leq c_{i,j}\\
										1,&z_{i,j}>c_{i,j}
										\end{array}\right.
\end{equation}
where 
$$z_{i,j}=\sqrt{n}\left(\frac{1}{2} \ln\left(\frac{1+r_{i,j}}{1-r_{i,j}}\right)-\frac{1}{2} \ln\left(\frac{1+\rho_0}{1-\rho_0}\right)\right)$$
 $c_{i,j}$ is $(1-\alpha_{i,j})$-quantile of standard normal distribution $N(0,1)$, $\alpha_{i,j}$ is the given significance level.

\begin{figure}[!h]
\includegraphics[scale=.25]{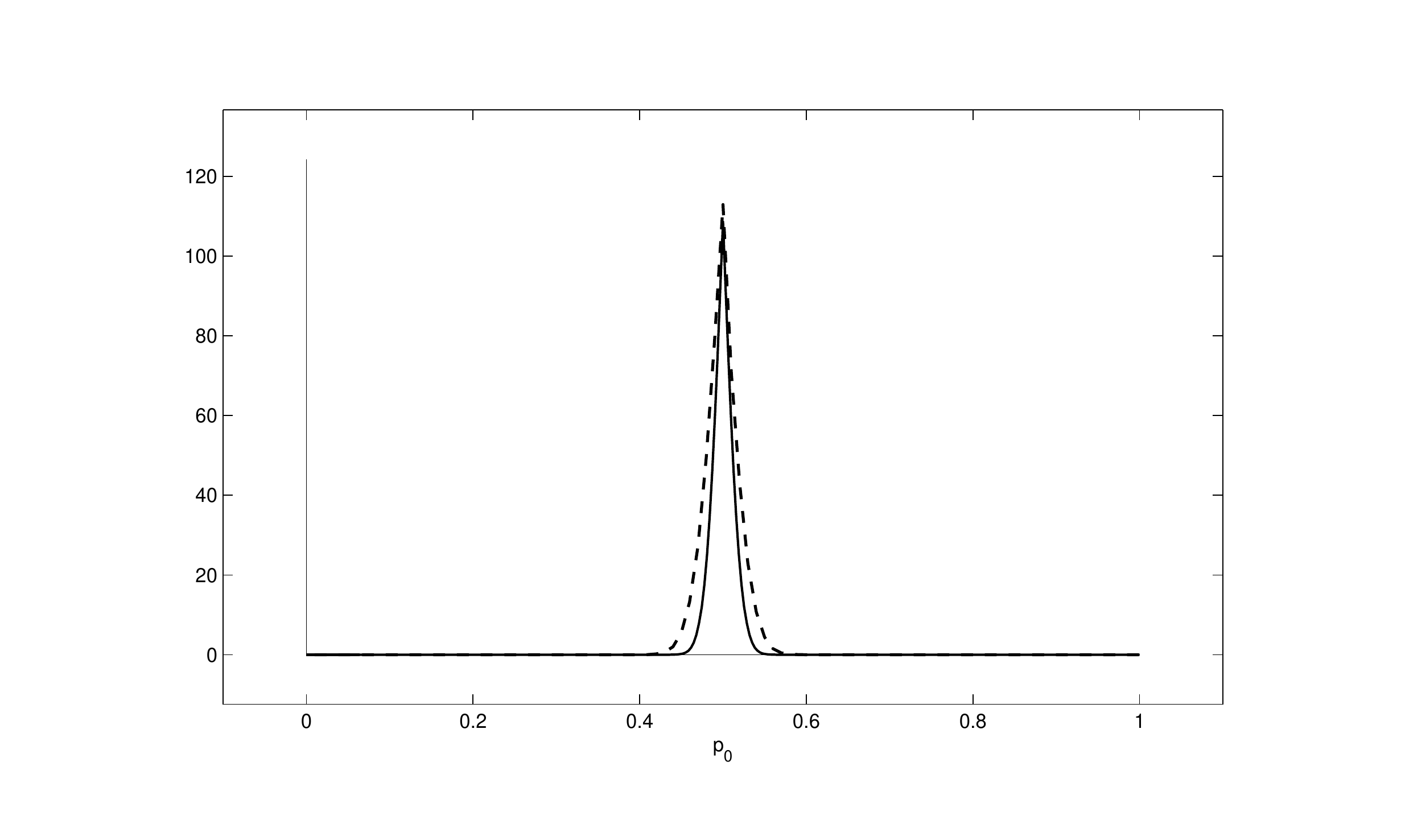}
\includegraphics[scale=.25]{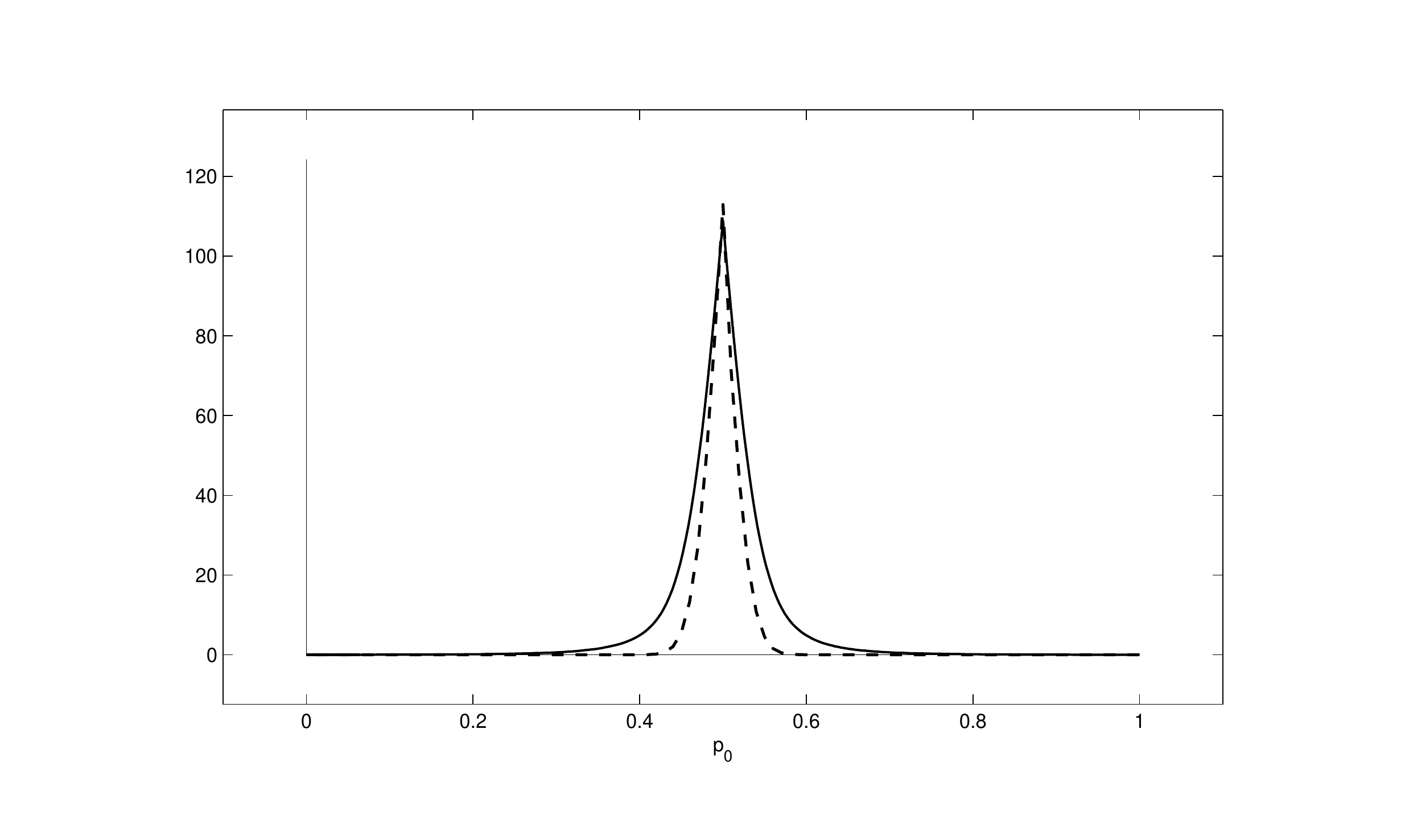}
\caption{Risk functions  $Risk(p_0)$ for the matrix $\Sigma_1$ and $\alpha=0,5$. Left: Gaussian distribution. Right: Student distribution. Solid line - Pearson correlation network. 
Dashed line - sign similarity network. 
Horizontal axe represents the value of $p_0$.}
\label{Matrix_1_alpha_0_5}
\end{figure}

Let $\Phi^{P}(x)$ be the matrix
\begin{equation}\label{test_for_N_hypotheses_Pearson}
\Phi^{P}(x)=\left(\begin{array}{cccc}
1,&\varphi^{P}_{12}(x),&\ldots,&\varphi^{P}_{1N}(x)\\
\varphi^{P}_{21}(x),&1,&\ldots,&\varphi^{P}_{2N}(x)\\
\ldots&\ldots&\ldots&\ldots\\
\varphi^{P}_{N1}(x),&\varphi^{P}_{N2}(x),&\ldots,&1\\
\end{array}\right).
\end{equation}
where $\varphi^{P}_{ij}(x)$ are defined by (\ref{Pearson_test}). 
Define the following multiple statistical procedure 
\begin{equation}\label{mdp_another_form_Pearson}
\delta^{P}(x)=d_G, \  \mbox{iff} \  \Phi^{P}(x)=G
\end{equation}
It is proved in \cite{KKKP13} that $\delta^P$ is optimal in Pearson correlation Gaussian network in the class of $w$-unbiased procedures, which have individual edge $(i,j)$ tests based on observations $x_i(t),x_j(t):i,j=1,\ldots,N$. 
Note, that both procedures $\delta^{Sg}$ and $\delta^P$ can be used for any distribution of vector $X$.

\begin{figure}[!h]
\includegraphics[scale=.25]{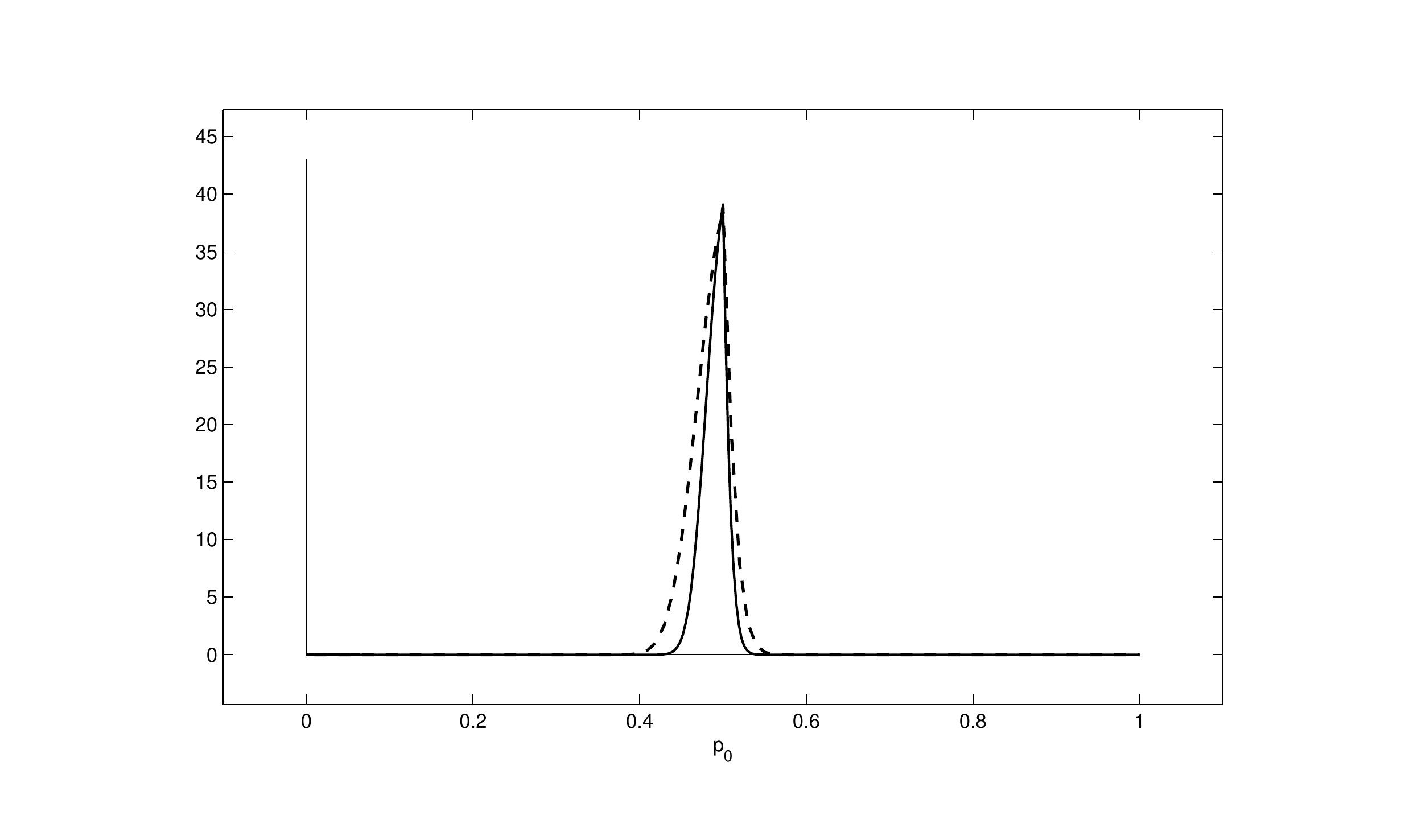}
\includegraphics[scale=.25]{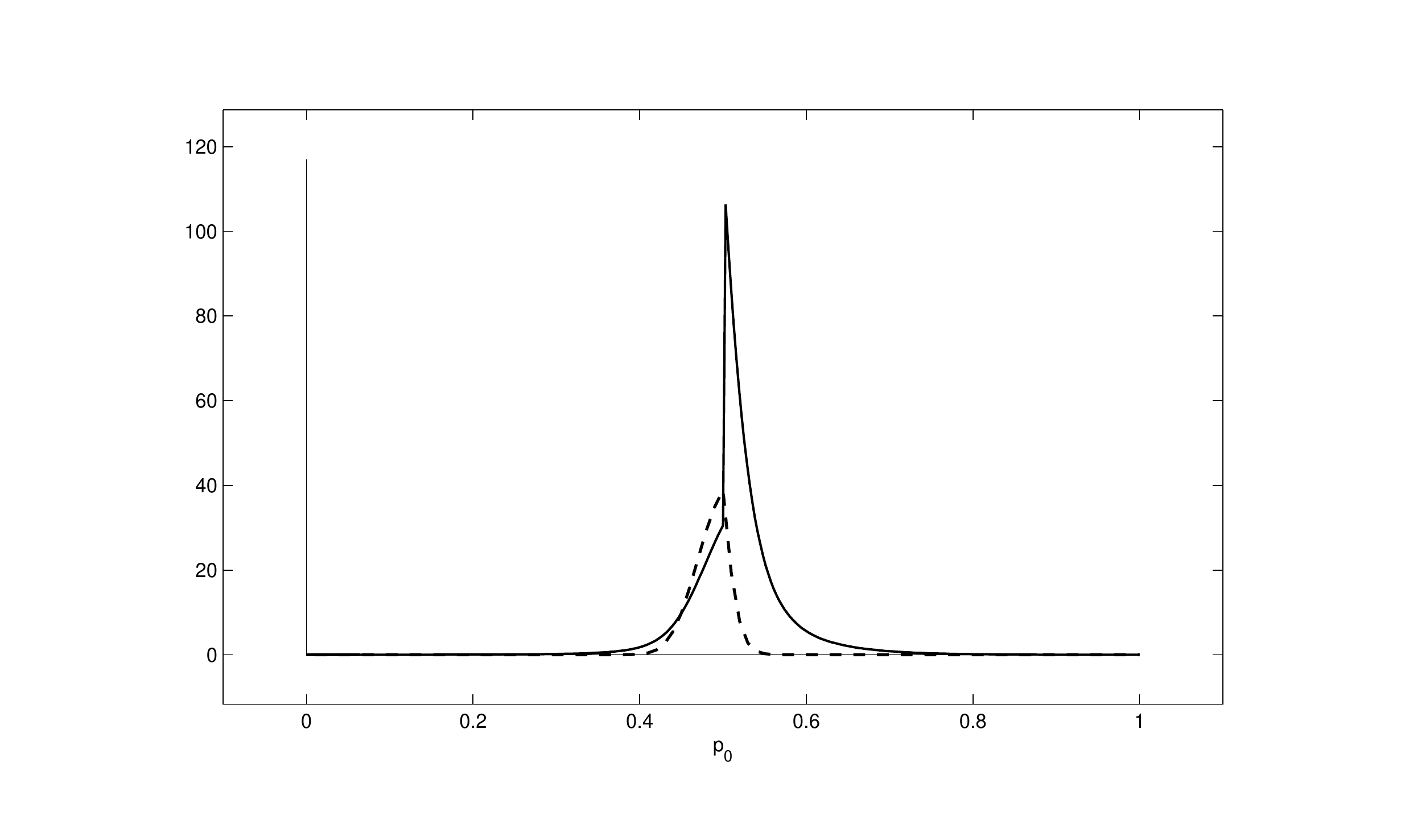}
\caption{Risk functions  $Risk(p_0)$ for the matrix $\Sigma_1$ and $\alpha=0,1$. Left: Gaussian distribution. Right: Student distribution. Solid line - Pearson correlation network. 
Dashed line - sign similarity network. 
Horizontal axe represents the value of $p_0$.}
\label{Matrix_1_alpha_0_1}
\end{figure}

\begin{figure}[!h]
\includegraphics[scale=.25]{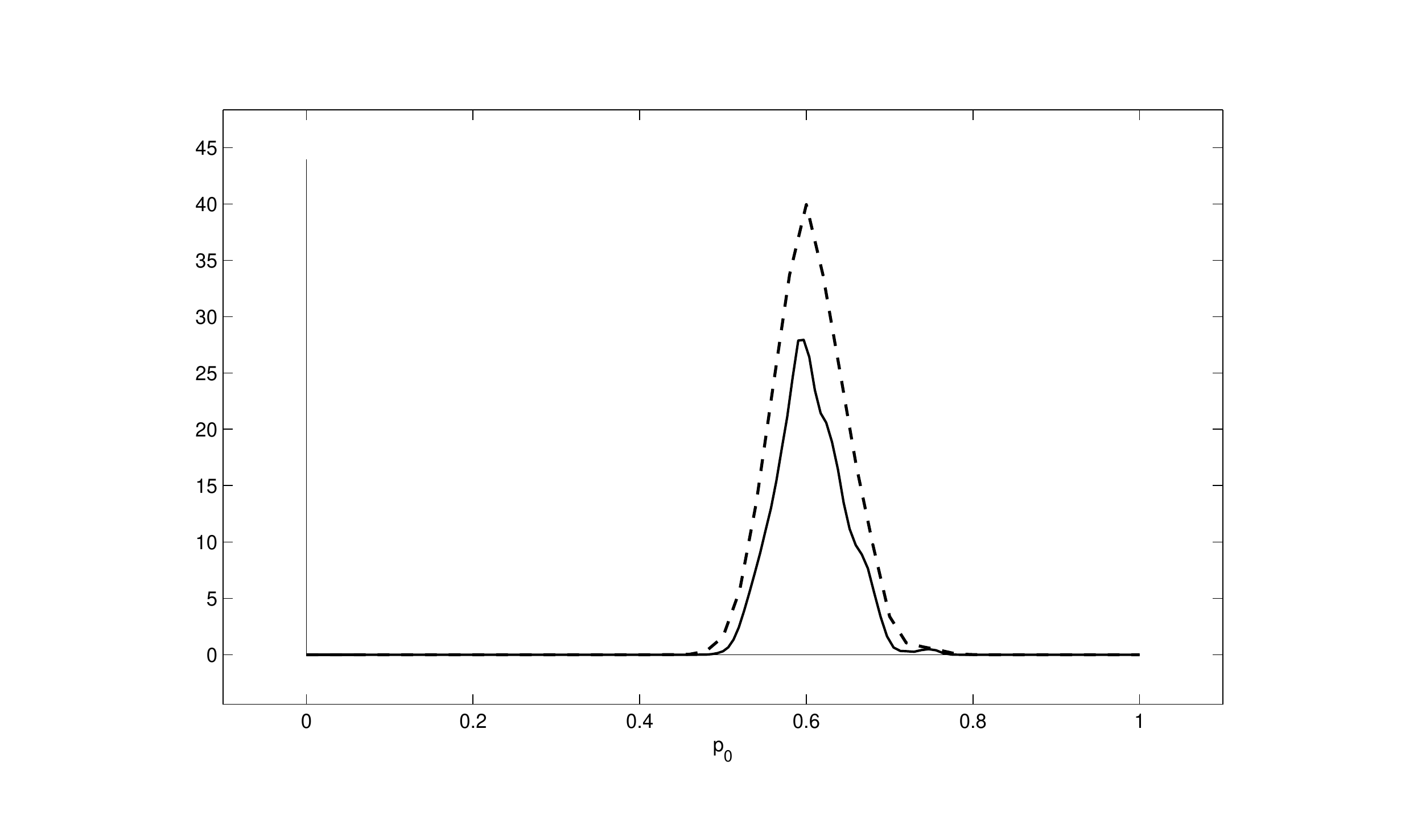}
\includegraphics[scale=.25]{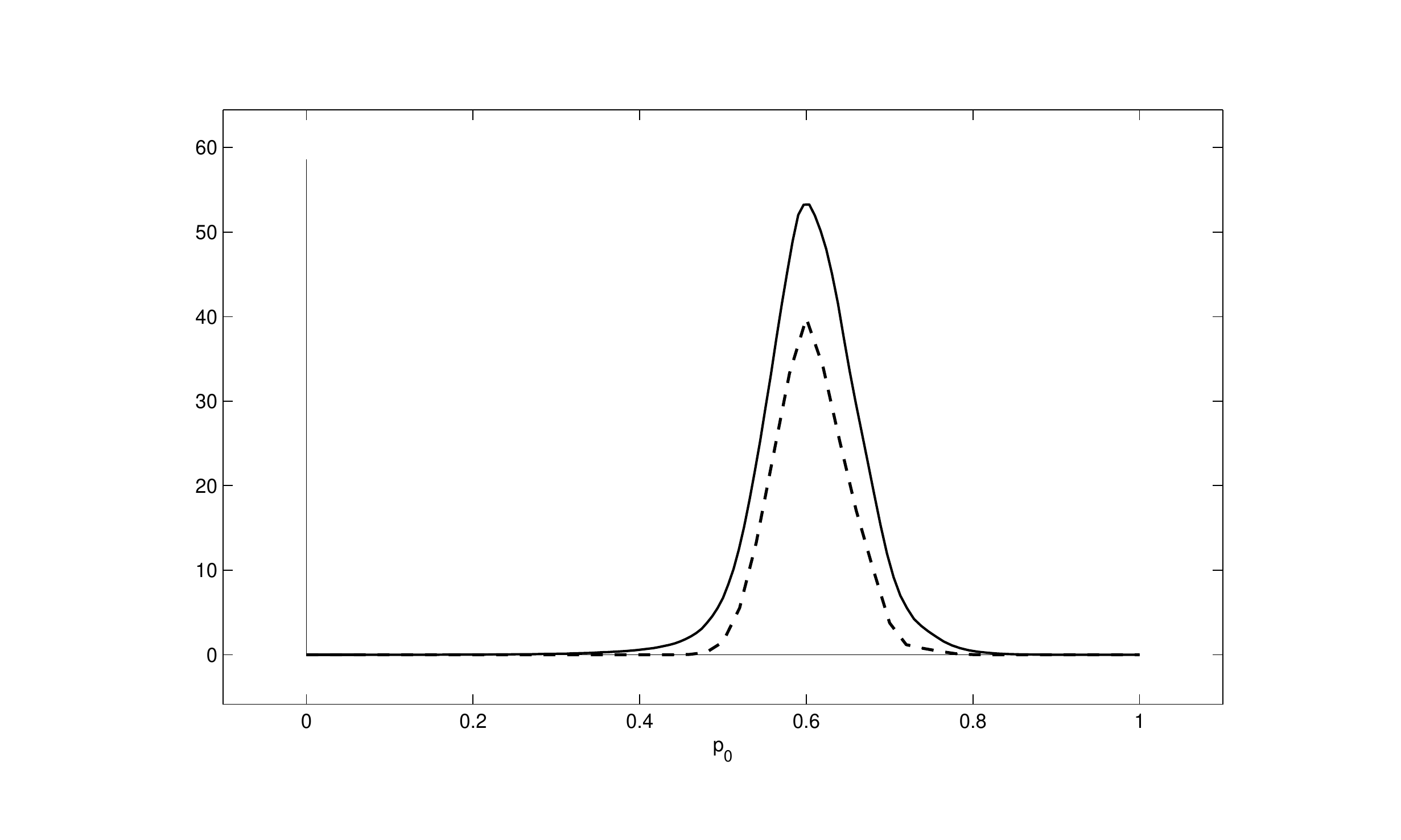}
\caption{Risk functions  $Risk(p_0)$ for the matrix $\Sigma_2$ and $\alpha=0,5$. Left: Gaussian distribution. Right: Student distribution. Solid line - Pearson correlation network. 
Dashed line - sign similarity network. 
Horizontal axe represents the value of $p_0$.}
\label{Matrix_2_alpha_0_5}
\end{figure}

To compare the risk functions we use two type of multivariate distributions (multivariate Gaussian and Student distributions) and 
three types of correlation matrices: 
\begin{enumerate}
\item Zero correlations matrix $\Sigma_1=\mbox{diag}(1,1,\ldots,1)$
\item Real correlations matrix $\Sigma_2$ calculated from stocks of Dow-Jones index of  USA market for 2013. 
\item High correlations matrix $\Sigma_3=(\sigma_{i,j})$, with $\sigma_{i,i}=1$, $\sigma_{i,j}=0,9$, $i \neq j$. 
\end{enumerate}
We choose the following values of parameters: $N=30$, $n=400$, significance level for all individual tests $\alpha=0,5$, $\alpha=0,1$. 

\begin{figure}[!h]
\includegraphics[scale=.25]{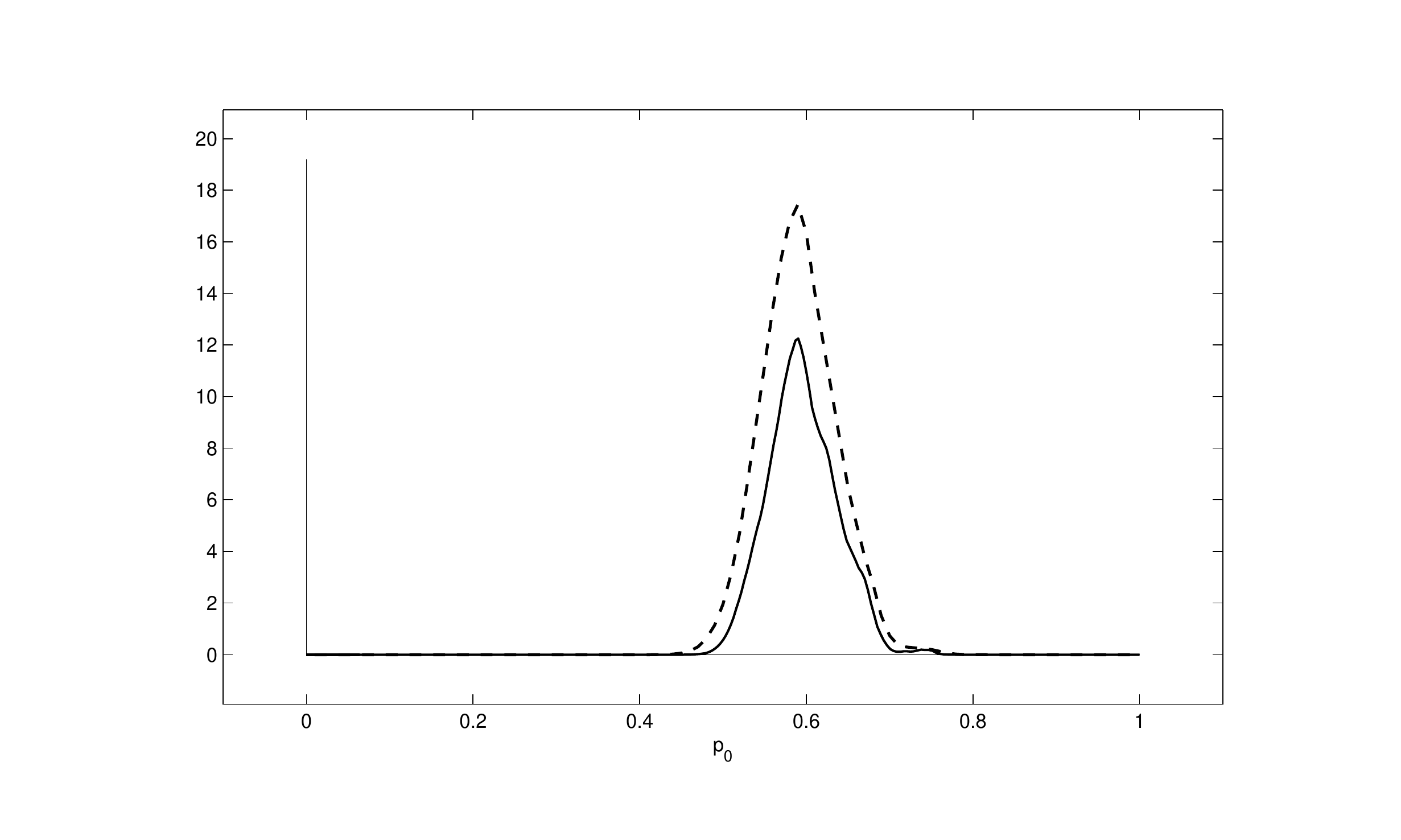}
\includegraphics[scale=.25]{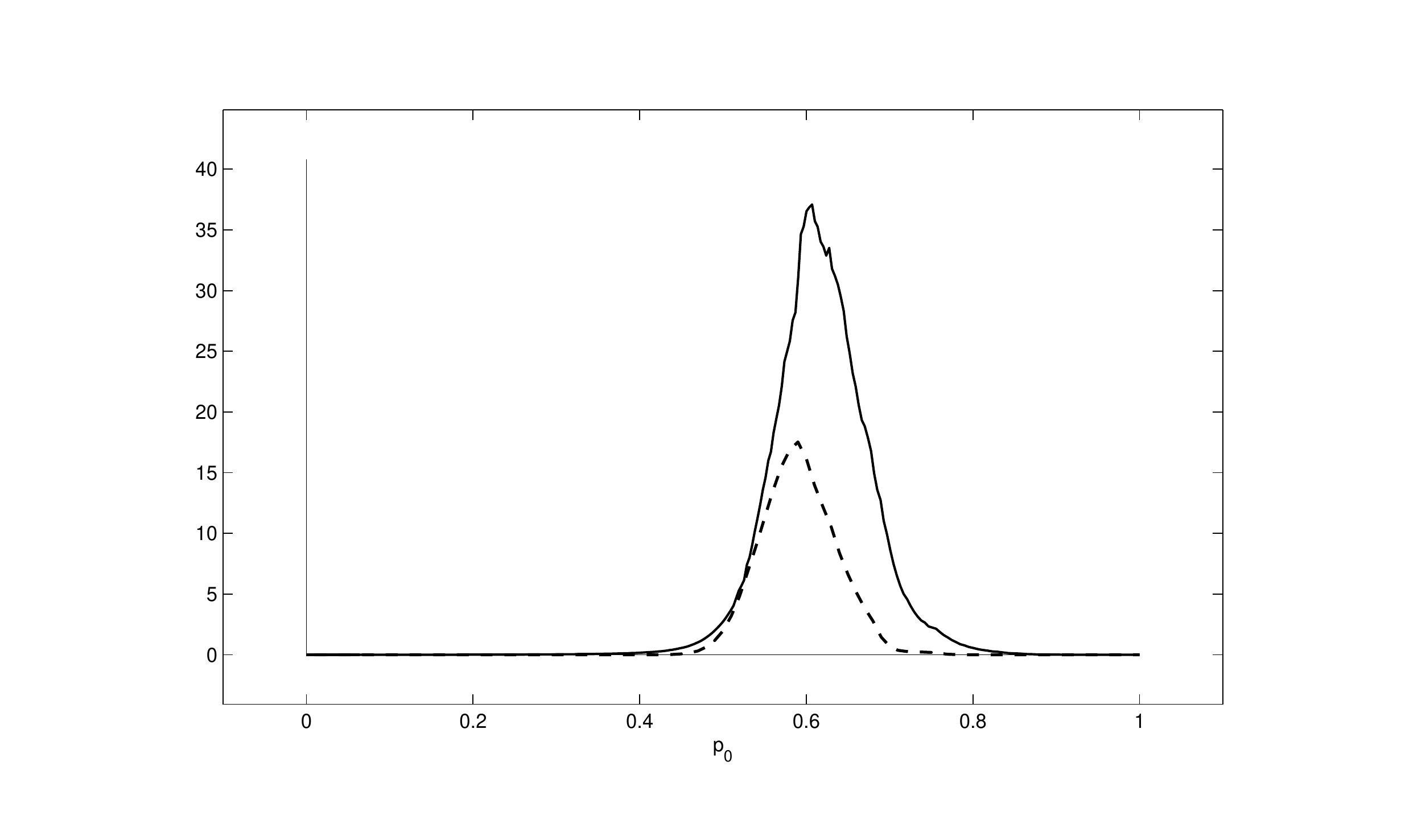}
\caption{Risk functions $Risk(p_0)$ for the matrix $\Sigma_2$ and $\alpha=0,1$. Left: Gaussian distribution. Right: Student distribution. Solid line - Pearson correlation network. 
Dashed line - sign similarity network. 
Horizontal axe represents the value of $p_0$.}
\label{Matrix_2_alpha_0_1}
\end{figure}

\begin{figure}[!h]
\includegraphics[scale=.25]{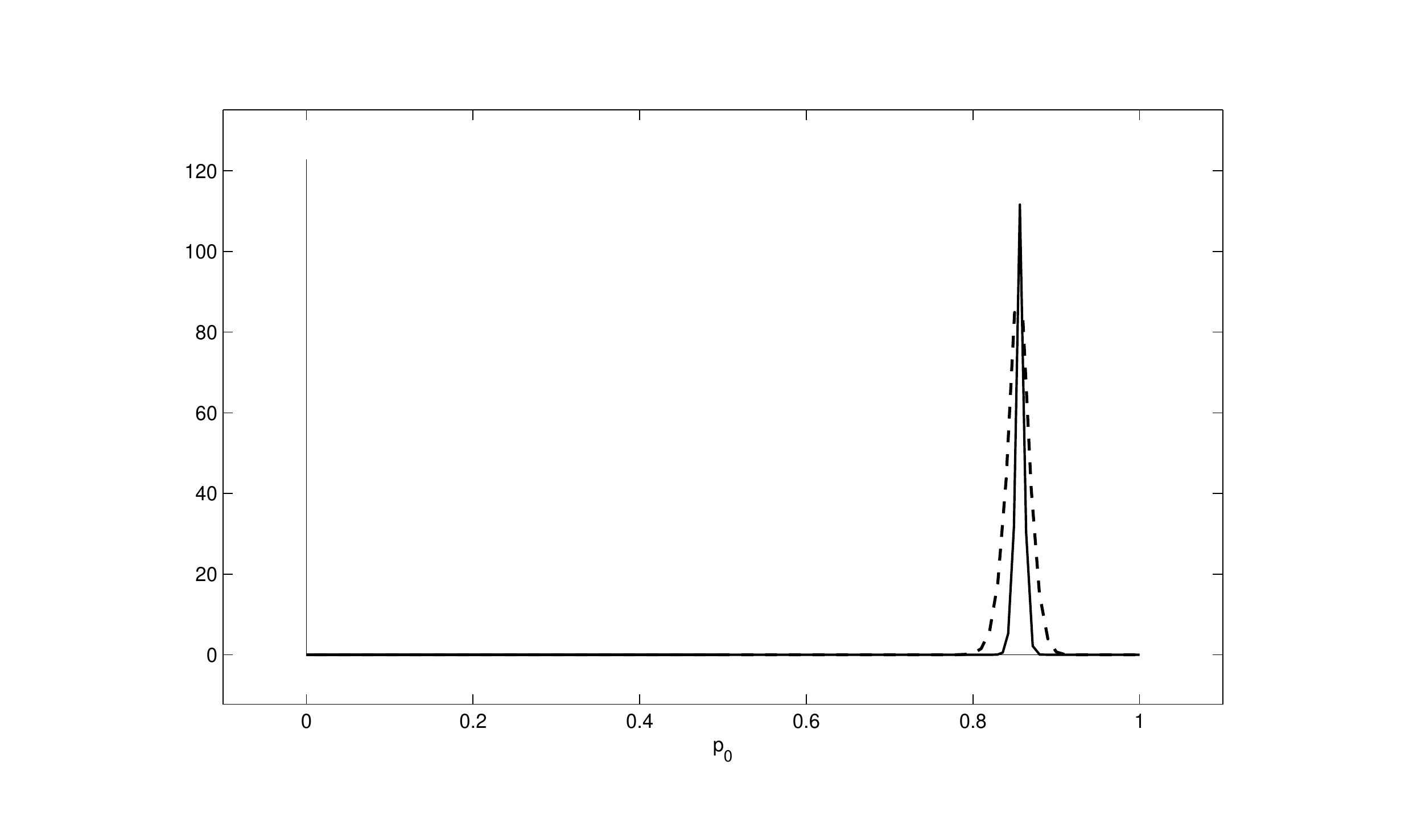}
\includegraphics[scale=.25]{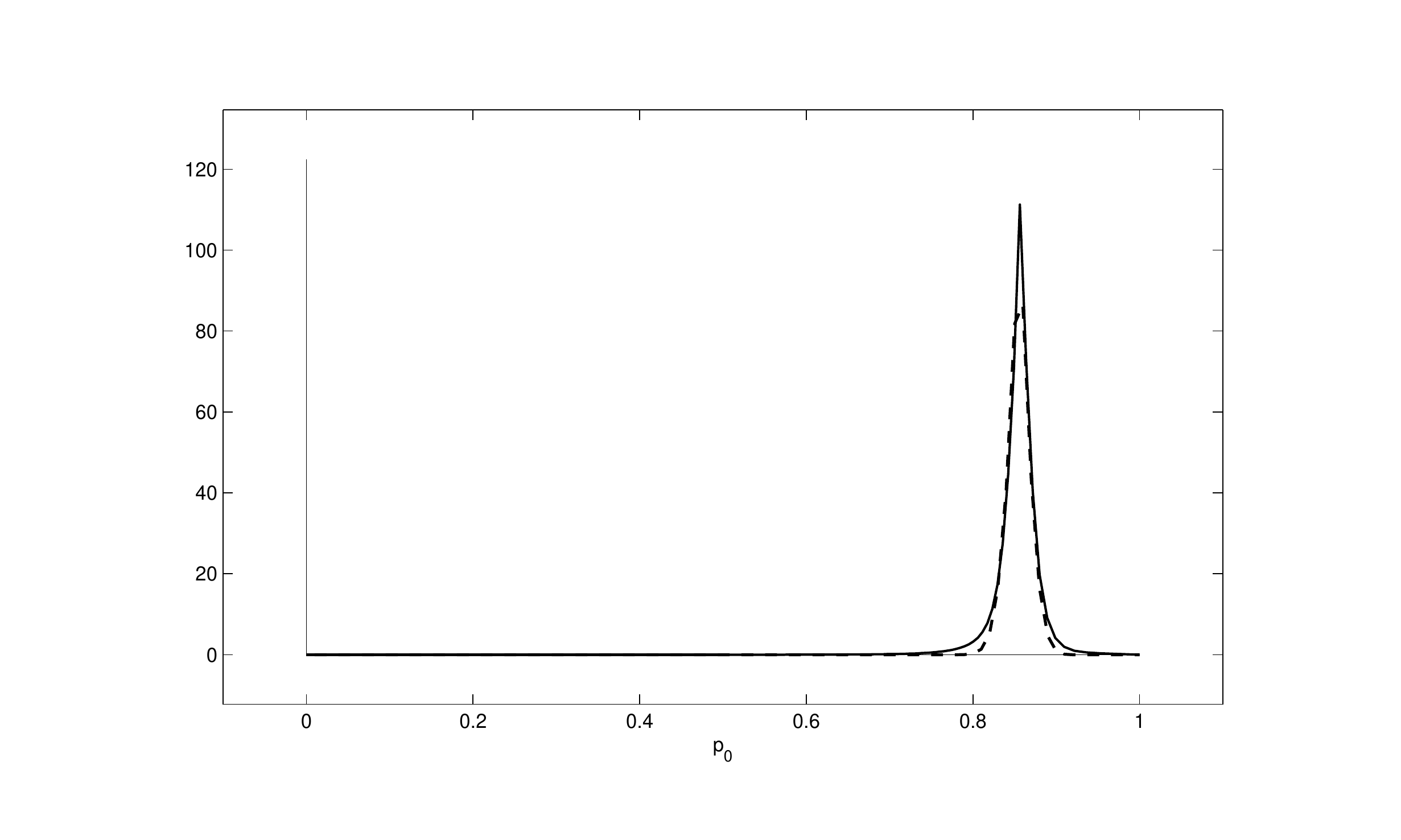}
\caption{Risk functions  $Risk(p_0)$ for the matrix $\Sigma_3$ and $\alpha=0,5$. Left: Gaussian distribution. Right: Student distribution. Solid line - Pearson correlation network. 
Dashed line - sign similarity network. 
Horizontal axe represents the value of $p_0$.}
\label{Matrix_3_alpha_0_5}
\end{figure}

We are interested in behavior of risk function as a function of threshold. For Pearson correlation network the value of threshold $\rho_0$ is taken from the interval $[-1,1]$. 
For sign similarity network the value of threshold $p_0$ is taken from the interval $[0,1]$. To make a correct comparison we use the following transformation formula from the Pearson correlation
to the probability of sign coincidence:
$$
p=\frac{1}{2}+\frac{1}{\pi} \mbox{arcsin}(\rho)
$$
This formula is known for Gaussian distribution (\cite{Kramer}, Ch.21), and it can be proved for Student distribution too.

\begin{figure}[!h]
\includegraphics[scale=.25]{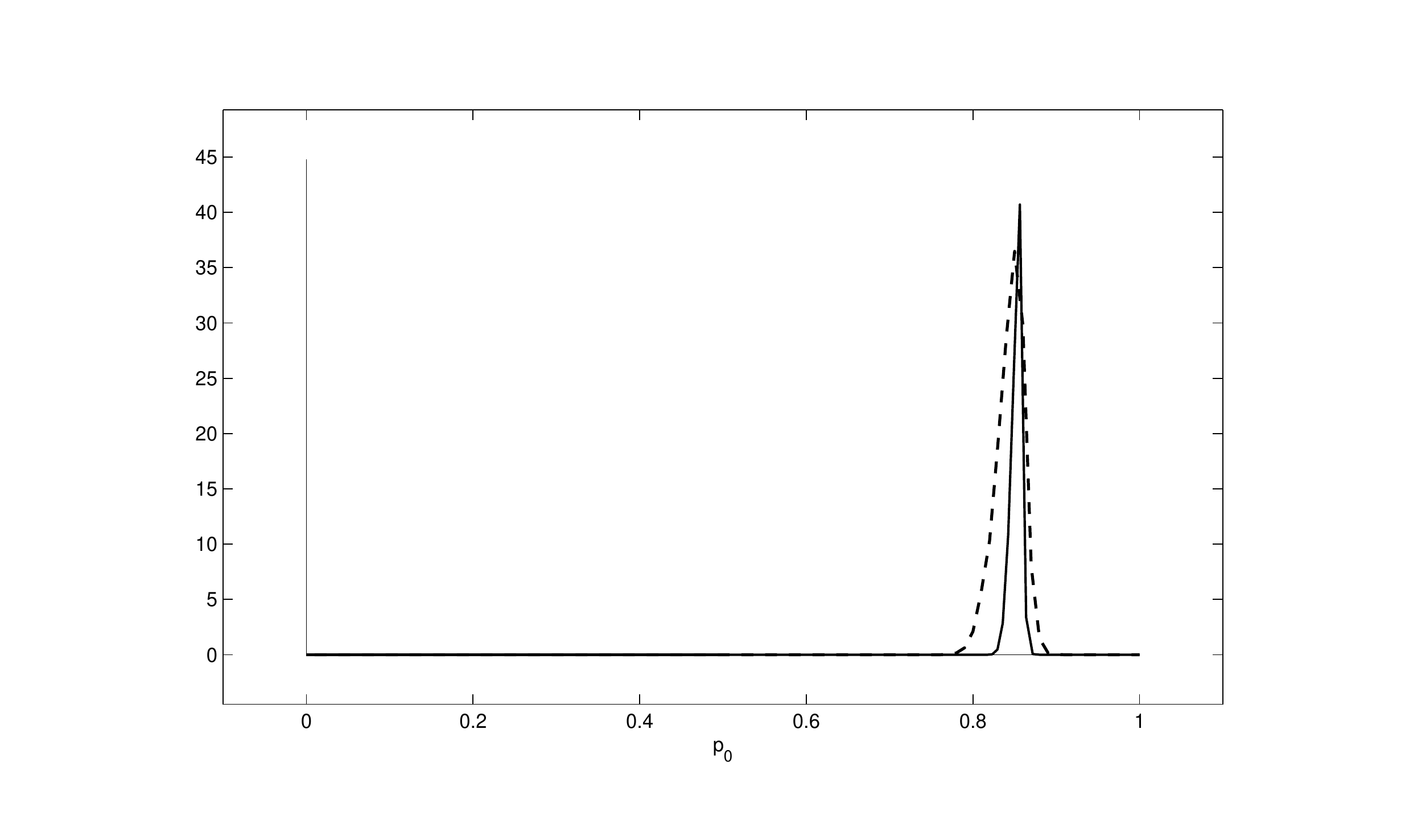}
\includegraphics[scale=.25]{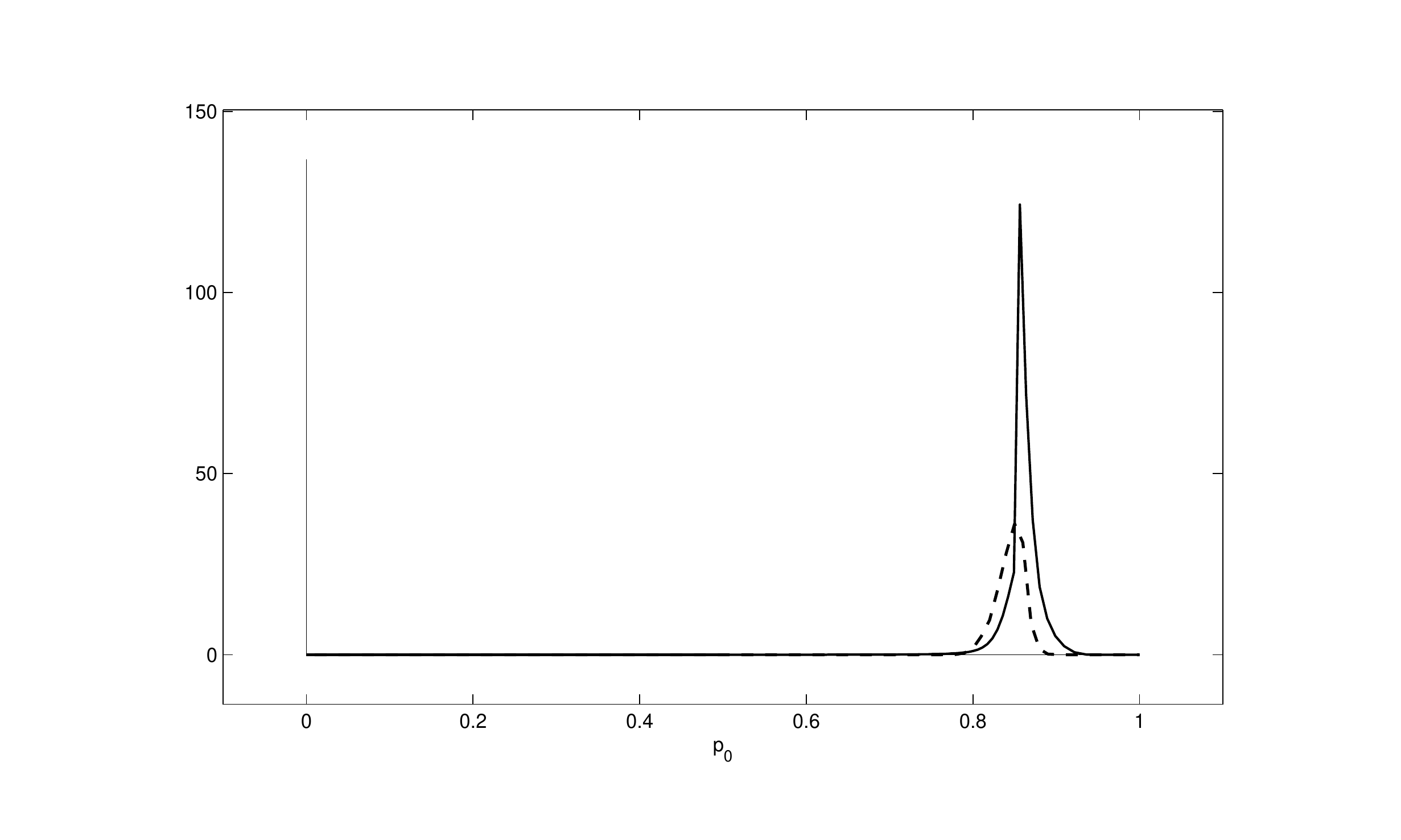}
\caption{Risk functions $Risk(p_0)$ for the matrix $\Sigma_3$ and $\alpha=0,1$. Left: Gaussian distribution. Right: Student distribution. Solid line - Pearson correlation network. 
Dashed line - sign similarity network. 
Horizontal axe represents the value of $p_0$.}
\label{Matrix_3_alpha_0_1}
\end{figure}

The results of numerical experiments are presented in Figures \ref{Matrix_1_alpha_0_5}-\ref{Matrix_3_alpha_0_1}. Figures \ref{Matrix_1_alpha_0_5}-\ref{Matrix_1_alpha_0_1} present the behavior of risk function of $\delta^{Sg}$ and $\delta^P$ as a function of threshold for 
the correlation matrix $\Sigma_1$, and $\alpha=0,5$ (Figure \ref{Matrix_1_alpha_0_5}), $\alpha=0,1$ (Figure \ref{Matrix_1_alpha_0_1}). One can see that both procedures control the risk function for Gaussian 
distribution. Namely, with the change of $\alpha$ from 0,5 to 0,1 the maximal value of risk function for both procedures is decreasing approximately from 120 to 40. In contrast, for Student distribution the maximal value of
risk function for the procedure $\delta^{Sg}$ is still decreasing, but the maximal value of risk function for the procedure $\delta^P$ keeps the same value.  This phenomenon is confirmed for the matrix $\Sigma_2$ by  comparison of Figure \ref{Matrix_2_alpha_0_5} and Figure \ref{Matrix_2_alpha_0_1} and for the matrix $\Sigma_3$ by  comparison of Figure \ref{Matrix_3_alpha_0_5} and Figure \ref{Matrix_3_alpha_0_1}. It means that the procedure $\delta^{Sg}$ controls the risk function for both distributions while the procedure $\delta^P$ does not. This gives advantage to the procedure $\delta^{Sg}$ for multivariate Student distributions for small value of $\alpha$.

\section{Concluding remarks}\label{Conclusions}
In this paper we introduce and investigate a class of statistical procedures with high reliability for the market graph identification in sign similarity  network.  
Theoretical investigation is conducted in the  framework of multiple decision theory.  Optimality of multiple decision procedure $\delta^{Sg}$ is proved under the following assumptions: 
additivity of loss functions,  unbiasedness of procedures, sign symmetry conditions and known expectations $E(X_i)$, $j=1,2,\ldots, N$. 
Additivity of the loss function and unbiasedness of procedures are appropriate for considered problems. Sign symmetry conditions are satisfied 
for a large class of distributions  used in financial analysis, in particular for elliptically contoured distributions. This class includes multivariate Gaussian and   Student distributions with heavy tails. 
Practical advantage of constructed procedures with respect to traditional ones is a high reliability of identification in a larger class of distributions.

Aknowledgement: The first author is supported by RFFI grant 14-01-00807, 
the second author is supported by RHF grant 15-32-01052, all authors are partly supported by  National Research University Higher School of Economics Scientific Found for International research laboratories.

\end{document}